# Comprehensive data-driven analysis of the impact of chemoinformatic structure on the genome-wide biological response profiles of cancer cells to 1159 drugs

Suleiman A. Khan[1,§], Ali Faisal[1], John Patric Mpindi[2], Juuso A. Parkkinen[1], Tuomo Kalliokoski[3], Antti Poso[4], Olli P. Kallioniemi[2], Krister Wennerberg[2], Samuel Kaski[1,5,§]

[1] Helsinki Institute for Information Technology HIIT, Department of Information and Computer Science, Aalto University, PO Box 15400, Espoo, 00076, Finland. [2] Institute for Molecular Medicine Finland, University of Helsinki, PO Box 20, Helsinki, 00014, Finland. [3] CADD, Global Discovery Chemistry, Novartis Institute for Biomedical Research, Basel, CH4002, Switzerland. [4] School of Pharmacy, Faculty of Health Sciences, University of Eastern Finland, PO Box 1627, Kuopio, 70211, Finland. [5] Helsinki Institute for Information Technology HIIT, Department of Computer Science, University of Helsinki, PO Box 68, Helsinki, 00014, Finland

[§]Corresponding authors. Contact: suleiman.khan@aalto.fi, samuel.kaski@hiit.fi

## Abstract

**Motivation:** Detailed and systematic understanding of the biological effects of millions of available compounds on living cells is a significant challenge. As most compounds impact multiple targets and pathways, traditional methods for analyzing structure-function relationships are not comprehensive enough. Therefore more advanced integrative models are needed for predicting biological effects elicited by specific chemical features. As a step towards creating such computational links we developed a data-driven chemical systems biology approach to comprehensively study the relationship of 76 structural 3D-descriptors (VolSurf, chemical space) of 1159 drugs with the gene expression responses (biological space) they elicited in three cancer cell lines. The analysis covering 11350 genes was based on data from the Connectivity Map. We decomposed these biological response profiles into components, each linked to a characteristic chemical descriptor profile.

**Results:** The integrated quantitative analysis of the chemical and biological spaces was more informative about protein-target based drug similarity than either dataset separately. We identified ten major components that link distinct VolSurf features across multiple compounds to specific biological activity types. For example, component 2 (hydrophobic properties) strongly links to DNA damage response, while component 3 (hydrogen bonding) connects to metabolic stress. Individual structural and biological features were often linked to one cell line only, such as leukemia cells (HL-60) specifically responding to cardiac glycosides.

**Conclusions:** In summary, our approach identified specific chemical structure properties shared across multiple drugs causing distinct biological responses. The decoding of such systematic chemical-biological relationships is necessary to build better models of drug effects, including unidentified types of molecular properties with strong biological effects.

## Background

The mechanism of action of drugs at the biochemical level has typically been studied by investigating specific chemical properties of the drug and the biological properties of its specific target [1,2]. This is the standard paradigm in Quantitative Structure Activity Relationship (QSAR) studies, where multivariate mathematical models are used for modeling the relationships between a set of physiochemical or structural properties and biological activity. In previous QSAR studies, such as in the classical 3D-QSAR work by Cramer et al. [3], values of a single biological activity measure are predicted.

However, biological responses at the cellular level are diverse and each drug typically binds to a multitude of targets in the cells and elicits a number of diverse effects. Systems-level approaches are thus needed to get a more comprehensive view of drug effects in living cells. Genome-wide massively multivariate description of the cellular responses caused by the drugs (Lamb et al., [4]) requires new kinds of tools for analysis and interpretation.

Chemical systems biology has emerged at the interface of systems biology and chemical biology with the goal of constructing a systems-level understanding of drug actions. Systematic analysis of a network of drug effects, *i.e.* network pharmacology, offers great opportunities for drug design in the future [5]. Chemical systems biology has also been used to predict drug side effects [6] as well as in other types of toxicological analysis [7].

Here, we take a complementary approach, by studying the impact of a host of chemical descriptors on the resulting detailed drug response profiles on a genome-wide scale. We link key structural components of the drug molecules, as defined by 3D VolSurf descriptors, with the consistent biological properties, as measured by comprehensive gene microarray expression profiles. We have developed a data-driven approach to analyze relationships between patterns in the specific chemical descriptors of the drugs on one hand, and patterns in the genome-wide biological expression responses on the other, as shown in Figure 1.

As biological response data we use the Connectivity Map (CMap, [4,8]), which consists of gene expression measurements from three cancer cell lines (MCF7, PC3 and HL60) treated with over a thousand different drug molecules (Figure 1.C). These data offer a unique view to the genome-wide responses of the cells to drug treatments and has been used to find new biological links *e.g.* between heat shock protein inhibitors, proteasome inhibitors, and topoisomerase inhibitors [8].

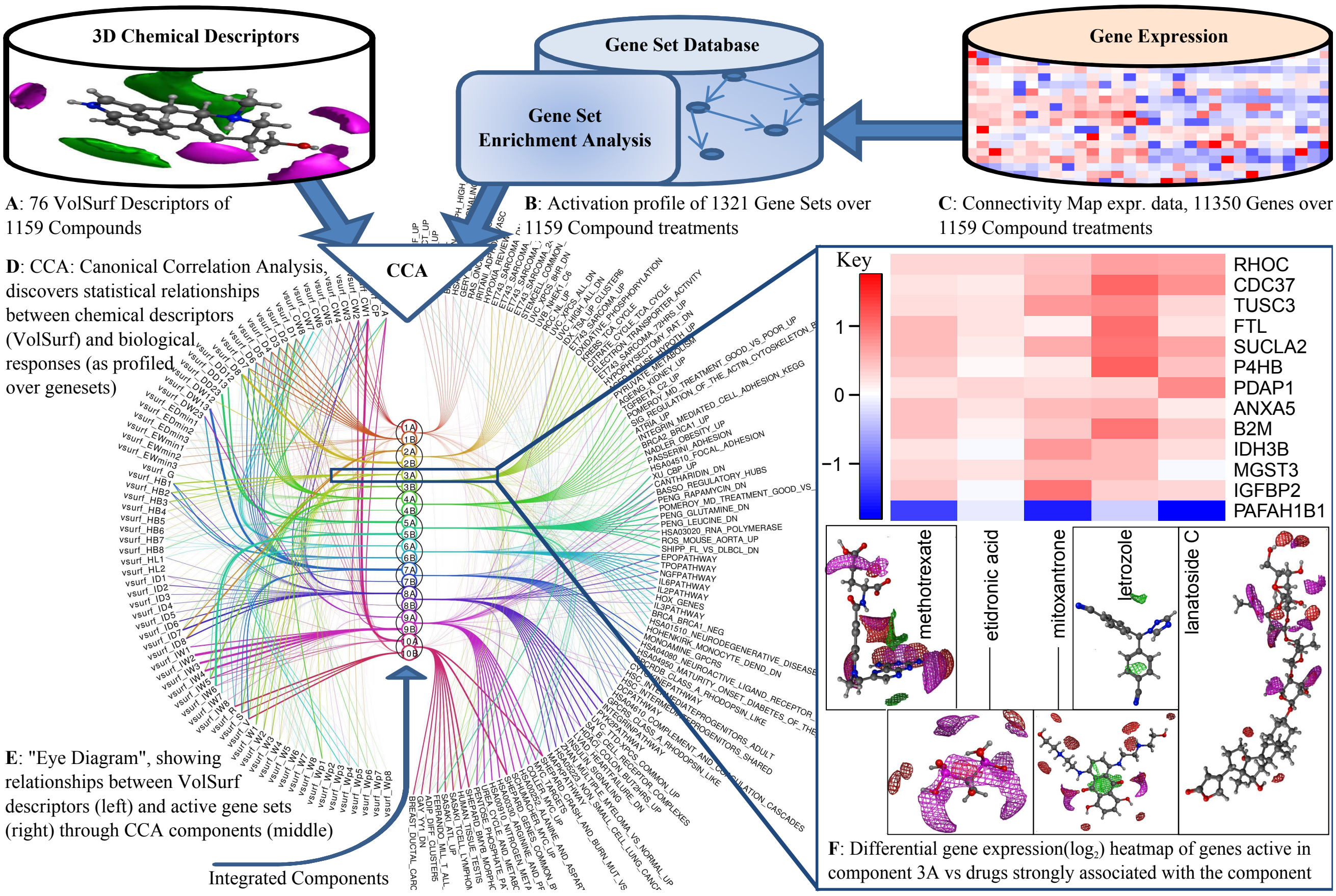

**Figure 1 - Data-driven search for statistical relationships between Chemical space (formed of VolSurf features) and Drug response space (gene expression).**

Our key assumption is that the chemical structure as encoded in the 3D descriptors of drugs impacts on the drug response resulting in specific patterns of gene expression. Furthermore, if there is any statistical relationship between the occurrence of patterns in the chemical space and the patterns in biological response space, those patterns are informative in forming hypotheses on the mechanisms of drug action. Given proper controls, the statistical responses can be attributed to the specific features of the chemicals tested out of a diverse drug library. In this paper we use comprehensive but readily interpretable models for finding the statistical dependencies. We search for distinct components that correlate the patterns in the chemical space with the biological response space. Assuming linear relationships, the task reduces to Canonical Correlation Analysis (CCA [9]) for searching for correlated components from the two data spaces (Figure 1.D). We visualize the components in a comprehensive way to facilitate interpretation (Figure 1.E and Figure 1.F) and validate them both qualitatively and quantitatively.

Canonical Correlation Analysis was recently used for drug side effect prediction and drug discovery by Atias and Sharan [10]. They applied CCA to combine known side effect associations of drugs with (i) 2D structure fingerprints and (ii) bioactivity profiles of the chemicals. The CCA results from both combinations were then successfully used to predict side effects for the drugs, suggesting that CCA is effective in finding relevant components from heterogeneous data sources.

Drugs generally act on a multitude of targets, several of which are connected to each phenotypic response. As most of this information is still unknown, modelling of the structure-target-response over a large drug library is not a straightforward goal. Therefore in this study we model the structure-response relationships of 1159 drug molecules directly, with CCA components playing the role of unknown mechanistic processes.

The non-availability of target information makes it important to select an appropriate type of chemical descriptors that allow capturing of generic response patterns. Many kinds of chemical descriptors are available for characterizing chemical structures in a quantitative way. Simple classical 2D fingerprints can be used to detect close analogs, but they would miss most if not all scaffold-hopping situations, where the different chemical entities give rise to similar pharmacophoric properties. Fingerprints and pure pharmacophoric descriptors require clearly defined individual targets, which are not known in many cases. In this piloting chemical and biological response study we use a set of VolSurf descriptors ([11]; Figure 1.A) that are ideal for capturing both structural similarities and general chemical features, such as solubility and permeation properties (ADMET: Absorption, Distribution, Metabolism, Excretion and Toxicology properties). Although VolSurf descriptors are not thought to explain detailed structure-activity relationships in the case of binding to single target, they offer a good overall interpretation of the molecular shape, hydrogen bonding, lipophilicity, and related properties, which are more conservative than individual binding cavities. It has also been shown that shape is a major factor when trying to find compounds with similar biological activity but dissimilar 2D structures [12].

The idea of correlating chemical structures with biological expression was introduced by Blower et al. in [13]. By combining 2D fingerprint data with biological activity profiles for the chemicals over 60 cancer cell lines (NCI60), and with steady-state gene expression measurements from those cell lines before drug treatments, they obtained indirect relationships between chemical substructures and the gene targets. In a more recent work, Cheng et al., [14] investigated correlations between the chemical structures, bioactivity profiles, and molecular targets for a set of 37 chemicals. This small-scale study demonstrated that combinations of biological activity and chemical structure information can provide insights into drug action mechanisms on a molecular level.

By using the direct gene expression responses to a large set of drug treatments from Connectivity Map, along with comprehensive component-level decomposition of response profiles, we are able to make more direct observations on how compounds impact on cells and what features of the chemical molecules and the biological responses are correlated with one another.

# Results and Discussion

We analysed the 1159 drug treatment gene expression responses of three cancer cell lines of the Connectivity Map, with the methods summarized in Figure 1 and detailed in Methods. The analysis decomposed into components the relationship between the "chemical space" of 76 chemical descriptors of each drug and the "biological space" of gene expression profiles of the drug response. We will call the components "CCA components" as the core method is Canonical Correlation Analysis (CCA). Each component relates a characteristic statistical pattern of gene expression with a pattern of the drug characteristics. In this section we analyse further the identified components and the statistical relationships they discovered.

## Quantitative validation of functional similarity of drug components

We evaluate the biological relevance of the extracted CCA components by studying the functional similarity of drugs associated with each component. In particular, we measure the performance of the component model to retrieve similar drugs, as indicated by external data about their function, and compare it to retrieval based on either the biological or chemical data separately. Details of the validation procedure are described in Methods. The mean average precision obtained for the retrieval task on the four data sets (CCA components, chemical space, biological space as represented by GSEA and Gene expression) are plotted in Figure 2.

The results show that retrieval based on the chemical space, *i.e.* VolSurf descriptors, performs clearly better than retrieval based on the biological space (activities of gene sets and genes), indicating that the chemical information is more relevant for evaluating the functional similarity of the chemicals. The biological space encoded by gene sets performs similarly to the original gene expression, indicating that the gene sets are a sensible encoding of the biological space; information lost due to dimensionality reduction is balanced by introduction of prior biological knowledge in the form of the sets. Finally, the combined space formed by the CCA components shows significantly better retrieval performance than either of the data spaces separately. The results are consistent over the range of the number of drugs considered in the retrieval task. These results show that CCA is able to extract and combine relevant information about the chemical structure and biological responses of the drugs, while filtering out biologically irrelevant structural information and also biological responses unrelated to the chemical characteristics.

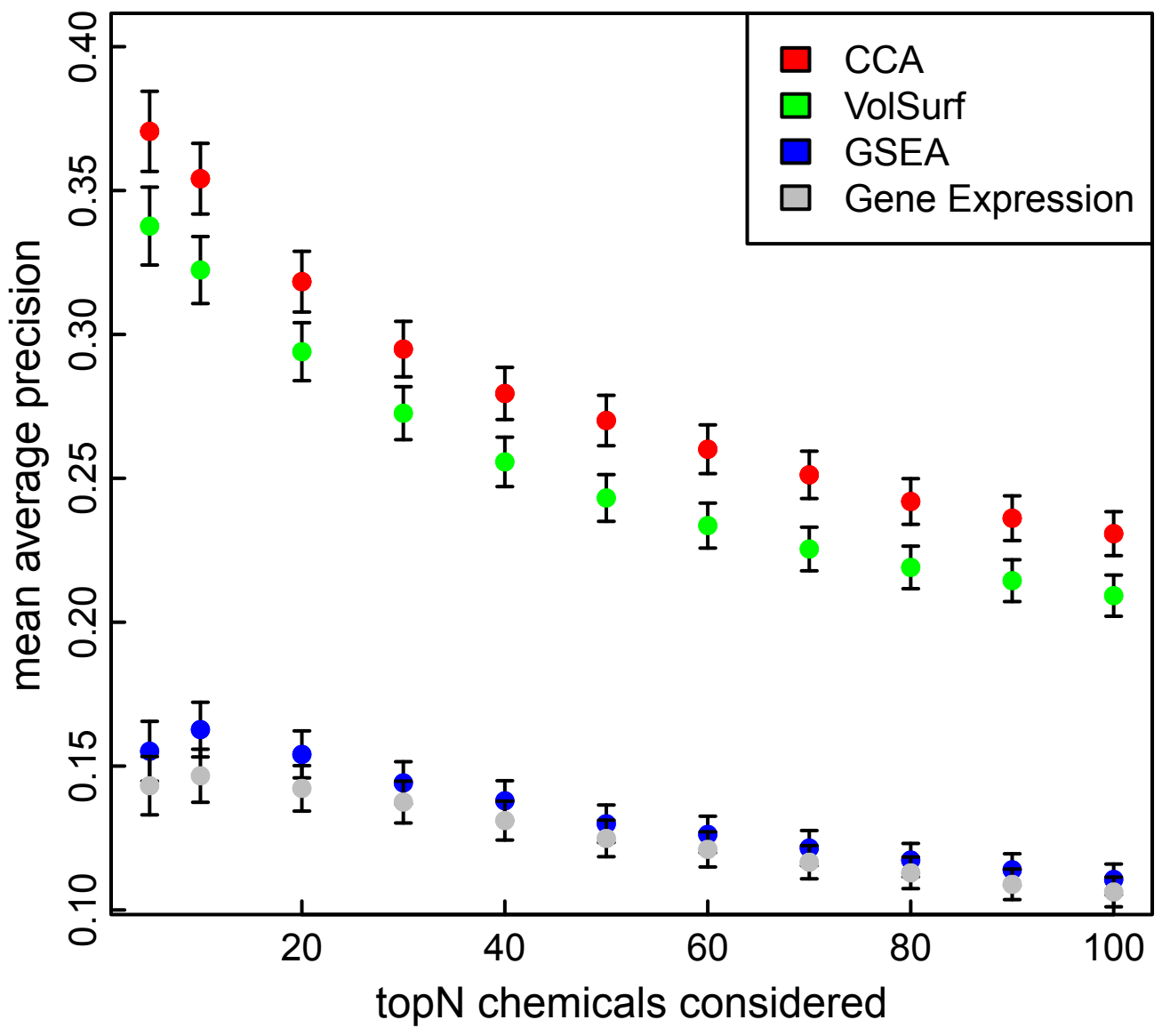

**Figure 2 - Quantitative validation of functional similarity of drug components.** The figure shows the mean average precision for retrieving functionally similar chemicals as a function of the number of top chemicals considered. Results are shown for three representations: CCA (red), Chemical space (green), and Biological space (GSEA: blue, Gene expression: grey). Error bars show one standard error of the mean precision.

## Response components and their interpretations

We next analyze the top ten CCA components having the highest significant correlations between the spaces. Figure 3 summarizes the relationships between the VolSurf descriptors and the gene sets as captured by the components. Each component is divided into two subcomponents 'A' and 'B', where in the first the compounds have positive canonical score and in the second negative, even though the characteristic response patterns are otherwise the same (details in Methods).

VolSurf descriptors, unlike more typically used 2D or 3D fingerprints and pharmacophores, do not have clear structural counterparts such as fragments or functional groups. Therefore, these descriptors are able to group together compounds with totally dissimilar chemical structures and yet having the same type of chemical properties. This is especially important in our study since the same biological activity may not have been created because of the same biological target protein but because of another target protein in the same pathway. Within a pathway, binding cavity properties may change, but general descriptors describing the size, lipophilicity, and shape are more conservative. Therefore VolSurf descriptors are better suited for classifying these aspects of the compounds.

In the case of the 1$^{st}$ and 2$^{nd}$ components, the selected VolSurf descriptors are dominated by hydrophobic-related descriptors while other descriptors are not present. Another component with major hydrophobic contribution is 4$^{th}$, but in this case also the molecular volume and surface area are having some effect. The 3rd component is connected to hydrogen bonding, polar interactions, and dispersion-related descriptors. Components 5-7 are mainly connected to "pharmacophoric" descriptors that are describing distribution of strong interaction points over the molecular space. Components 8 and 9 are both strongly influenced by integy-moments, describing imbalance of either hydrophobic or hydrophilic areas over the whole molecular volume. Component 10 is mainly affected by shape and size-related parameters, and also lipophilic integy moments.

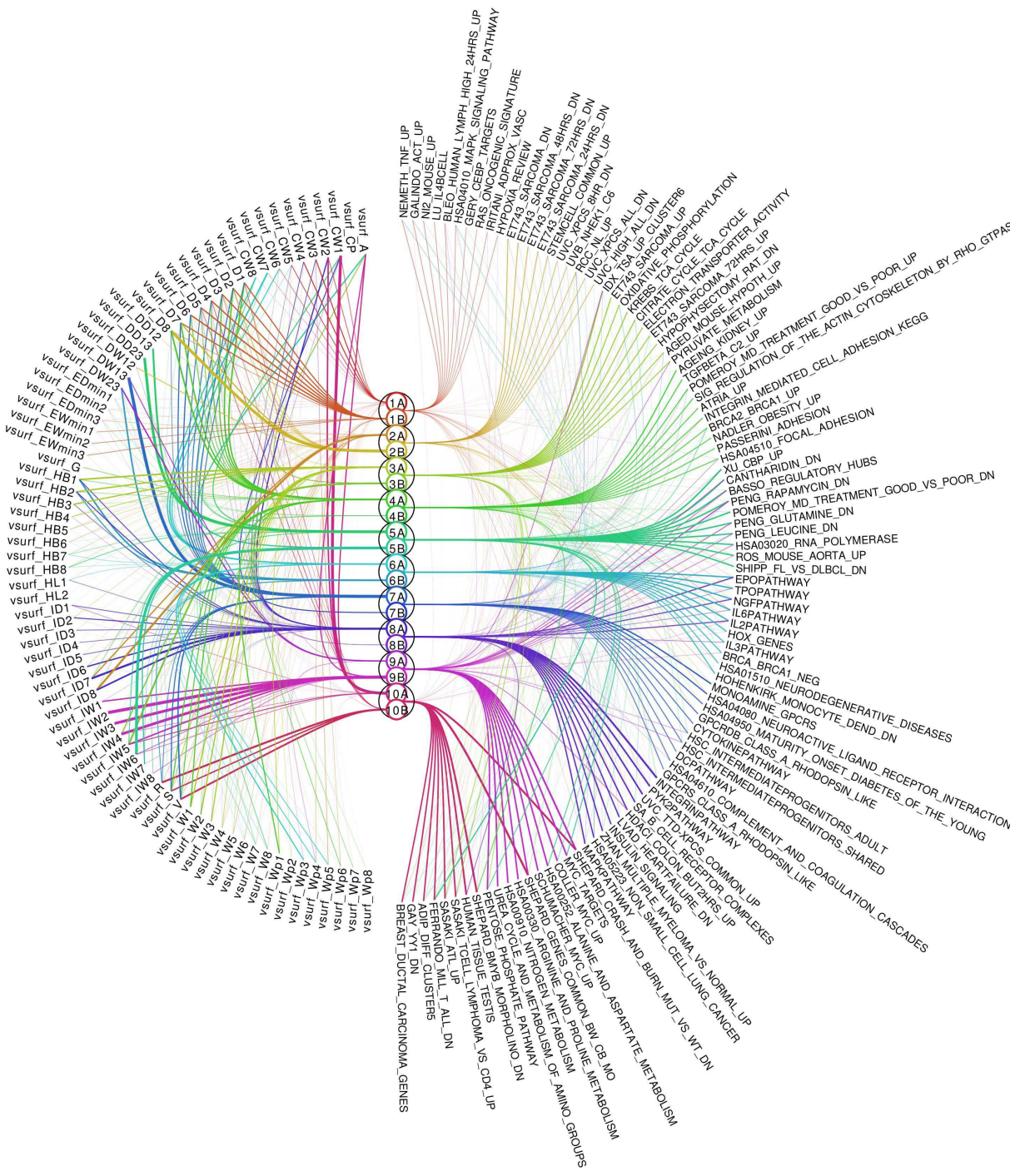

**Figure 3 - Relationships decomposed into components.** "Eye diagram" showing the top 10 significant CCA components ordered by correlation from top to bottom (middle), VolSurf descriptors (left), and top gene sets (right). The CCA components are shown as circles, with numbers indicating the decreasing order of canonical correlation and letters A and B indicating subcomponents (A: positive canonical score, B: negative canonical score). The widths of the curves from the components to VolSurf descriptors and gene sets indicate the strength of the corresponding associations. For VolSurf descriptors the subcomponent-specific activity is shown, whereas for the gene sets we show the overall activity in the component. (Zoom for details)

On the biological side of the components we observe that the enriched gene sets in component 1 indicate a mitogenic signaling response. Component 8 appears similar but has an additional link to cell adhesion signaling. Component 4 in turn very directly connects to cytoskeletal regulation and cell adhesion. While there appears to be a considerable overlap between compounds in components 2, 3 and 10, the enriched gene sets in component 2 show a strong link to DNA damage response, 10 is associated with common cancer signals, and component 3 is associated with an anabolic response. Components 5 and 6 are connected to different differentiation events. Component 7 links to gene expression changes seen in GPCR signaling. Component 9 links to amino acid and nitrogen metabolism.

We further extracted the significant genes in each component and performed GO enrichment analysis on them.

Based on the Eye diagram (Figure 3) and lists of significant genes, gene sets, GO terms, and the top 20 compounds, we summarize the biological and chemical patterns in Table 1.

## Component examples: 3/3A – A cell stress component

We observed that in component 3, the top genes and gene sets indicate mostly mitochondrial and metabolic stress-related processes. Top gene sets associating with this component include many gene sets connecting to mitochondrial function (Figure 3). Similarly, on the gene level several known cell stress genes such as PGK1, PGD, and PRMT1 [15-17] are upregulated. A deeper look into the 3D structures of the top compounds in this component reveals a possibility of 4-12 hydrogen bonds in all of the top compounds of set 3A. Thus, these compounds may be able to affect proton transportation processes, which is in agreement with the biological interpretation that mitochondria act like proton pumps. As an example, Figure 4 shows the hydrogen bond donor and acceptor regions of azacitidine and idarubicin.

**Table 1 - Summarized interpretation of top 10 components.** The pharmacophoric enrichment analysis (marked with "*") was carried out over VolSurf features considered as a gold standard, and measuring enrichment of the list in a component by a hypergeometric test.

| Comp. | Biological Interpretation | Compounds in Group A | Compounds in Group B | VolSurf Interpretation |
|---|---|---|---|---|
| 1 | Classic growth factor signaling: (MAP and protein kinase signaling) | Sulfonamides, antibiotics, carbonic anhydrase inhibitors | Antipsychotic and antihistaminic compounds | High lipophilicity |
| 2 | DNA damage | Contrast agents, antibiotics, | DNA damaging agents, antimetabolites | Strong lipophilic areas emphasized |
| 3 | Stress response, mitochondrial and anabolic metabolism | DNA damaging agents | GPCR antagonists, ion channel blockers | Polar interactions enriched |
| 4 | Cytoskeleton, cell adhesion and migration | GPCR liganda, macrocyclic cmpds and contrast agents | Beta adrenergic agonists, other GPCR ligands | N/A |
| 5 | Differentiation, EMT, stemness | NSAIDS, cAMP signaling promoting compounds | HDAC Inhibitors, HDAC-like | Significantly enriched with pharmacophoric features* |
| 6 | Inflammatory and differentiation signaling | N/A | Protein synthesis inhibitors, anti-diabetics, cardiac glycosides | Pharmacophoric features* |
| 7 | GPCR and cytokine signaling | N/A | Cardiac glycosides, cephalosporins | Pharmacophoric features* |
| 8 | Growth factor and cell adhesion signaling | Cardiac glycosides | β-adrenergic agonists, $Ca^{2+}$ channel blockers | Integy-moment and significant pharmacophoric enriched* |
| 9 | Amino acid and nitrogen metabolism | Protein synthesis inhibitors | Anti-diabetics | Integy-moment and significant pharmacophoric enriched* |
| 10 | Cancer signaling | DNA damaging agents | Corticosteroids, ionophores | Size shape type descriptors |

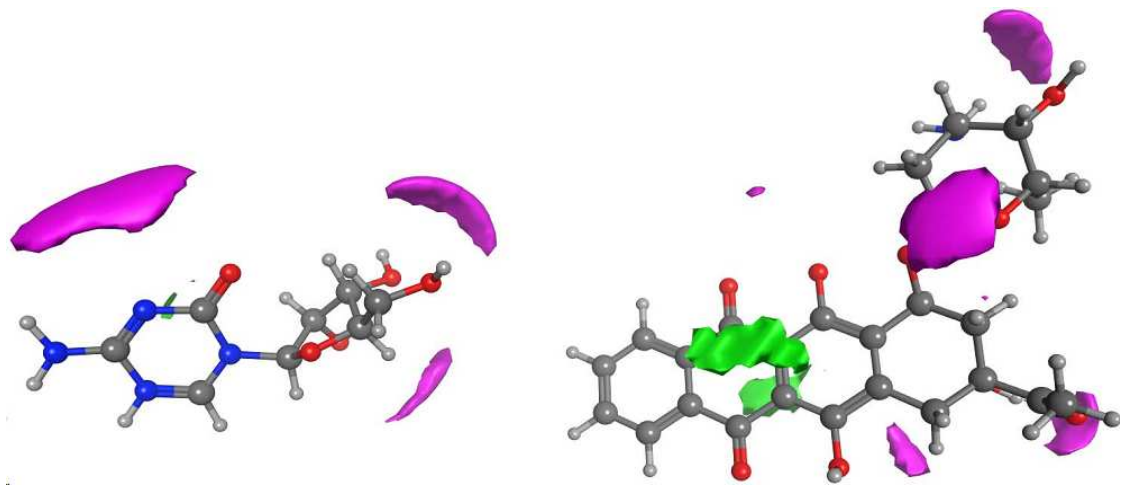

**Figure 4: Compounds high in hydrogen Bonding.** Azacitidine (left) and Idarubicin (right) showing H-bonding areas with blue (hydrogen-bond donor) and red (hydrogen-bond acceptor).

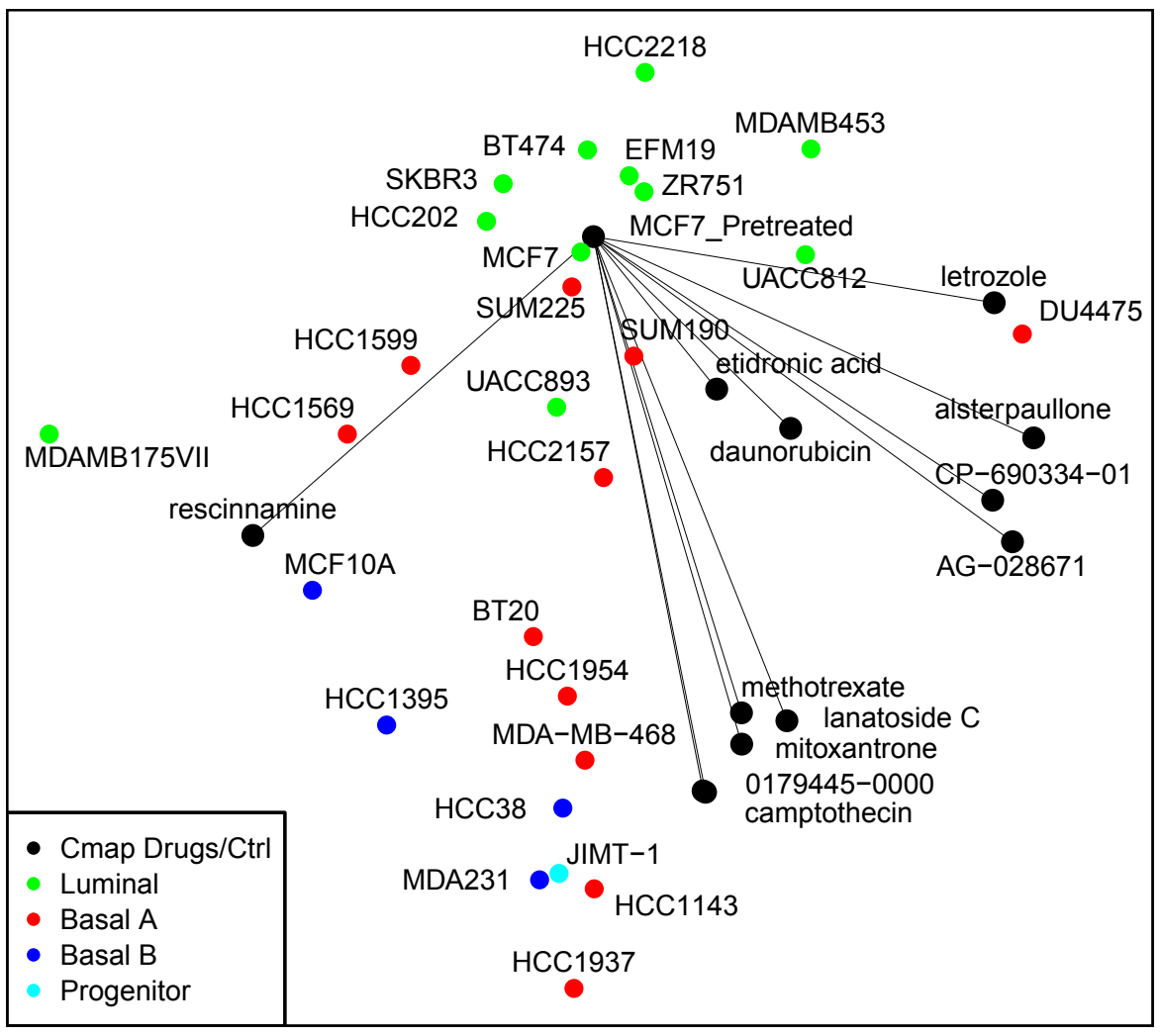

**Figure 5 - 3A drug transitions.** NeRV visualization showing Drug Treatment Transitions. Lines indicate the transition from Pretreated MCF7 to treated MCF7 cells.

To help understand how the biological variability caused by the drugs and the chemical features in component 3A compares with the intrinsic variability from one cell type to another, we visualize drug response transitions on MCF7 cells. Thirty different breast cancer cell types are used as references in their unperturbed condition (as described in Methods). The MCF7 treatments from subcomponent 3A and the thirty independent breast cancer cells are plotted in Figure 5 using a recent multidimensional scaling method called NeRV [29,30]. NeRV shows these different cell instances mapped onto the 2-dimensional display such that similarities are preserved as faithfully as possible. Subcomponent 3A contains many DNA-damaging agents such as the DNA intercalating and topoisomerase inhibitory camptothecin, daunorubicin, and mitoxantrone, the CDK inhibitor alsterpaullone, the cardiac glycoside lanatoside C, which at high concentrations is likely to inhibit topoisomerases [18] the antimetabolite methotrexate, as well as rescinnamine, which has been suggested to induce a DNA damage response without itself inducing DNA damage [19] and the aromatase inhibitor letrozole. The NeRV plot based on the top changed genes in treated and untreated MCF7 cells as well as a panel of other breast cancer cell lines, shows that after treatment with these drugs, the gene expression of the luminal, ER-positive MCF7 cells starts to resemble the basal, ER-negative breast cancer types. Interestingly, while MCF7 cells are relatively chromosomally stable, the drug-treatments make them appear like chromosomally unstable and intrinsic high DNA damage cell lines such as HCC1937 or MDA-MB-231. Cell stress is an emerging cancer therapy target and it is interesting that a) this subcomponent including topoisomerase inhibitors, antimetabolites and CDK inhibitors induces stress-related metabolic responses in MCF7 cells similar to what is seen in a basal level in other, more malignant and undifferentiated breast cancer cells, and b) it raises the hypothesis that the compounds in this subcomponent could be combined with cell stress targeting compounds. This finding is strengthened by the fact that many of the top upregulated genes in the 3A subcomponent; ACHY, CDC37, GPI, ME2, PMRT1, P4HB, PGD, and PGK1 are also overexpressed in breast cancers as compared to normal tissue.

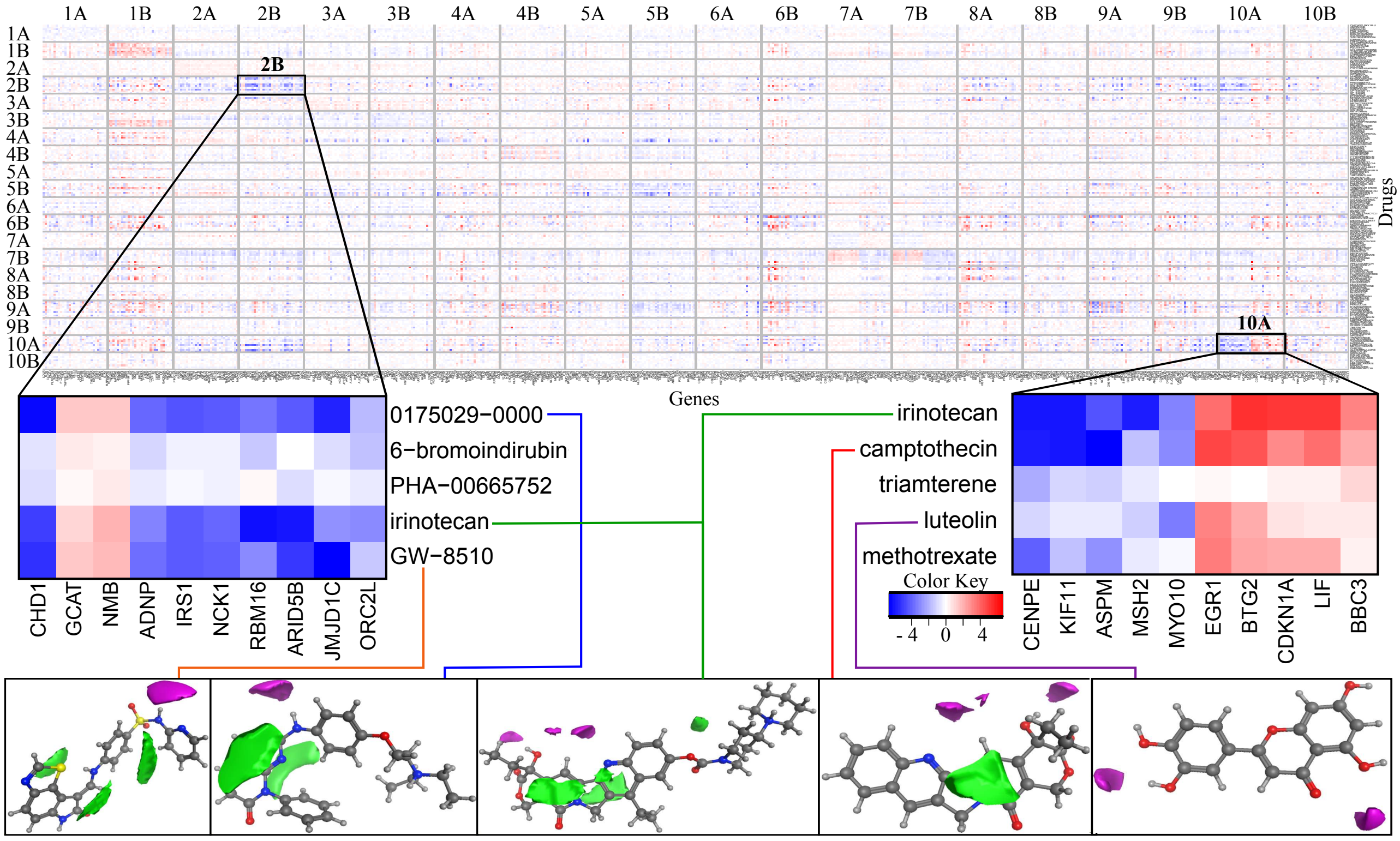

**Figure 6 - Finding interesting components.** Heatmap across the 10 highest scoring significant CCA components: X-axis lists the top 30 significant genes in each subcomponent, while y-axis represents the top 20 scoring compounds in each. Two unique components 2B and 10A are zoomed in to show the detailed expression pattern along with 3D VolSurf descriptors (green areas are the lipophilic fields and the purple water fields). Only a subset (5 compounds and 10 genes) is shown in the zoomed version.

## 2B & 10A – Functionally similar but gene-wise different responses

We observe that component groups 2B, 3A, and 10A share several compounds such as the DNA-intercalating topoisomerase inhibitors mitoxantrone and irinotecan, the cyclin-dependent kinase (CDK) inhibitors alsterpaullone and GW-8510, and the antimetabolites methotrexate and azacitidine, 5 of the top 20 between each paired group. Most of the non-overlapping compounds in each component group are not linked functionally or structurally in any obvious way, on the other hand. To verify that the components capture different phenomena despite sharing several compounds, we compute chemical composition and biological similarity matrices over all component pairs. We use the Tanimoto similarity measure to compute overlap between the top 30 genes of each subcomponent pair. The analysis of biological similarity between these subcomponents with compound overlap (out of top 20 compounds for each component) indicates that there is minimal biological and chemical sharing between any two components. Almost all component pairs that are highly biologically similar have a non-significant and low chemical composition similarity, and vice versa. This is a strong indication that we have identified sets of VolSurf descriptors that link to different biological responses. In some cases, several of these features can be identified in a single molecule like the etidronic acid, which is linked to both components 3 and 6. The chemical properties of component 6 are connected with pharmacophoric features and component 3 with hydrogen bonding, while biologically the components are related to differentiation and stress response, respectively.

To get a deeper view of the underlying biological response mechanisms we explore the response patterns of the components using heatmaps. In the first heatmap, we consider the most active genes in a subcomponent and plot their expression levels across the top compounds of every subcomponent (Figure 6 ). In the figure we search for the subcomponents that have a unique expression pattern across other subcomponents in a column. Components 2B and 10A show a unique structure. These seem to represent two separate aspects of DNA damage response, which are connected to two separate molecular features; hydrophobicity in component 2B and shape-type VolSurf descriptors in component 10A. The gene expression changes in both subcomponents are strongly linked to a DNA damage and mitotic arrest response with many proto-oncogenic cell division and mitogenic signaling genes being down regulated (Figure 6 ). The same genes are commonly seen upregulated in cancers and many of them have been and are pursued as drug targets. Therefore both the components are similar on a larger biological scale, but do in fact have little gene-wise overlap.

To validate these hypotheses, we checked for reported growth inhibition for the top 20 chemicals in these two subcomponents in the NCI/DTP *in vitro* cell line testing database (NCI60 testing program, http://dtp.nci.nih.gov/docs/cancer/cancer_data.html). Four compounds from 2B and 10 from 10A were represented in the NCI60 datasets (Table 2). For almost all of the compounds for which NCI60 data are available, in CMap the compounds were used at doses that very effectively will stop the cells from growing or kill them.

## 7B – A leukemia-specific subcomponent

Based on studying the heatmaps, 7B is another interesting subcomponent: It has a dominant effect on HL60 as compared to MCF7 and PC3, indicating that this subcomponent and the link

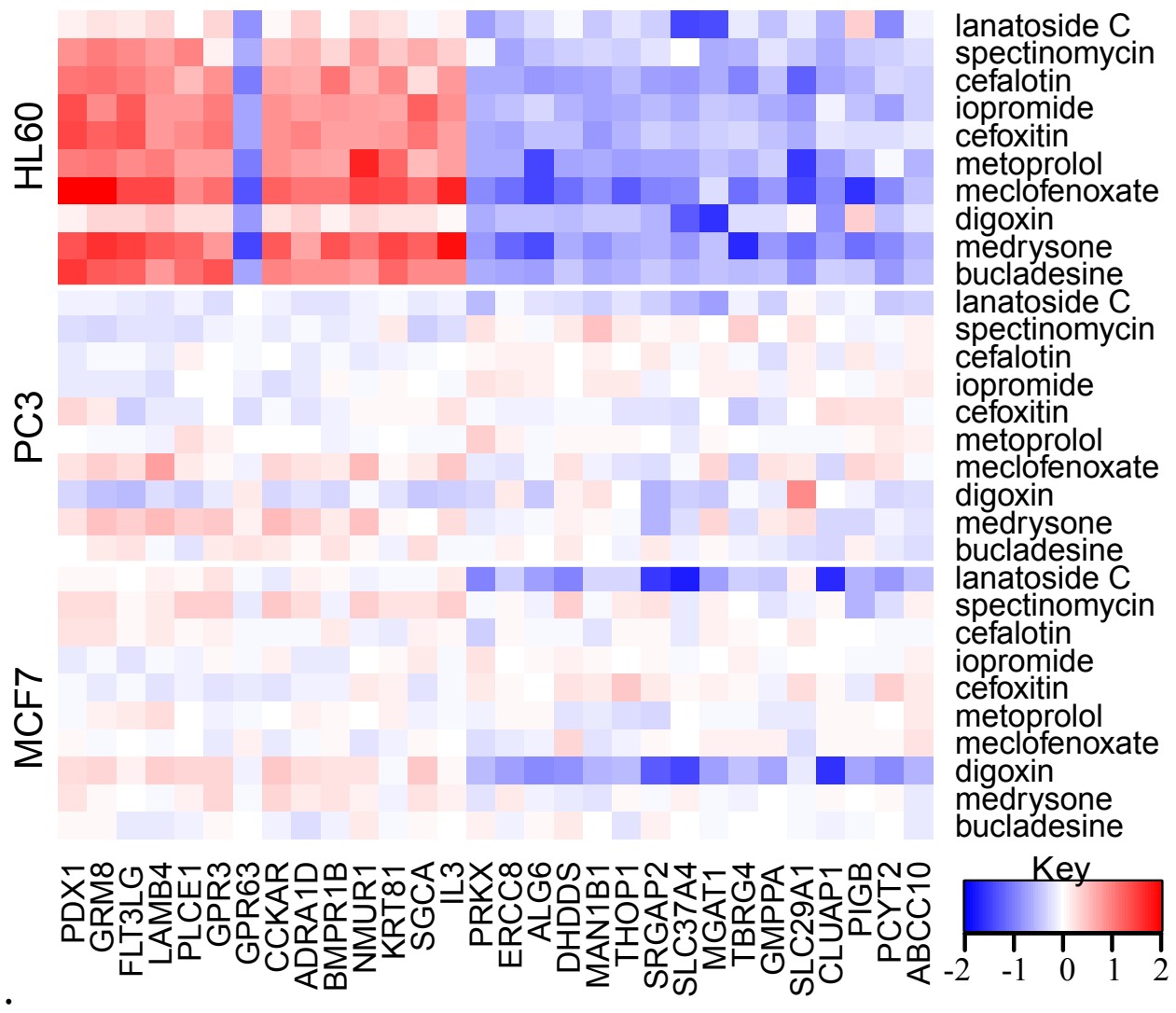

**Figure 7 - Heatmap for subcomponent 7B.** Y-axis lists the top 10 active compounds in the component, replicated over the three cell lines, while the X-axis lists the most significantly active genes in the component. The genes are clearly activated systematically and exclusively in the HL60 cell line, hence indicating an HL60 specific response.

**Table 2: Growth Inhibition verification of 2B/10A Compounds.** GI50 values (drug concentration causing a 50% growth inhibition) from NCI/DTP are shown along with the corresponding concentrations used in the Connectivity Map (CMap) data. By comparing the GI50 and CMap values we can get an idea of expected cell killing effect of the drug in the CMap data. Drugs that are expected to eventually kill the cells are shown in bold. Concentration values are in µM scales.

| Chemical | CMap (µM) | GI50 (µM) | Subcomponent | Cell line |
|---|---|---|---|---|
| **berberine** | **10** | **25.1** | 2B | MCF7 |
| **irinotecan** | **100** | **6.3** | 2B, 10A | MCF7 |
| **mitoxantrone** | **7.9** | **0.004** | 2B | MCF7 |
| **amiodarone** | **6.3** | **4.0*** | 2B | PC3 |
| **8-azaguanine** | **25.1** | **0.32** | 10A | HL60 |
| **apigenin** | **15.9** | **25.1** | 10A | HL60 |
| **azacitidine** | **15.9** | **0.79** | 10A | PC3 |
| **camptothecin** | **12.6** | **<0.01**** | 10A | MCF7 |
| **chrysin** | **15.9** | **15.8***** | 10A | MCF7 |
| **methotrexate** | **7.9** | **0.03** | 10A | MCF7 |
| **thioguanosine** | **12.6** | **0.32** | 10A | MCF7 |
| esculetin | 25.1 | >100** | 10A | HL60 |
| fulvestrant | 1.0 | >100** | 10A | PC3 |

* GI50 value at the end of the tested range
** Mean of GI50 values from HL60 and MCF7 cell lines
*** Value from HL60 cell line

between structure and gene expression may be specific for leukemic cells and leukemic-specific molecular targets.

Figure **7** shows the activity of most significant genes (columns) corresponding to the top compounds (rows) across the three cell lines. The top drugs are mainly cardiac glycosides and these drugs are known to have a strong toxic effect on leukemic cells at the concentrations used. It is worth noting that FLT3LG (FLT3 ligand) is one of the most significantly up-regulated genes. The FLT3 receptor, to which FLT3LG binds, is an emerging target in leukemia [20].

# Conclusions

We have introduced a chemical systems biology approach for analyzing the complex relationship patterns between chemical structures of drug molecules and their genome-wide responses in cells. With Canonical Correlation Analysis, we are able to find statistical dependencies between the two data spaces in the form of correlated components. We have demonstrated quantitatively that these components are more informative about drug similarity than either chemical or biological data separately.

In our study, we used gene sets to introduce biological knowledge into the analysis. Iorio et al. [8] have recently got very promising results with an alternative method of analyzing gene expression responses. It is an interesting and non-straightforward research question whether that approach can be generalized to searching for structure-response relationships.

We have also demonstrated the use of advanced visualization methods to facilitate detailed interpretation of the chemical and biological characteristics of the components. Our findings show connections between the biological responses of many known drug groups to their general chemical properties (Table 1). As an example of the ability of the model to discover detailed drug response mechanisms we were able to separate different DNA damage responses that appear to be driven by different chemical features in compound sets having considerable overlap. Subcomponents 2B, 3A, and 10A all contain classic chemotherapeutic and DNA damaging agents as described in the results section. Strikingly, subcomponents 2B and 10A are driven primarily by hydrophobic/lipophilic descriptors and are more similar in their biological output. They both connect to the downregulation of many proto-oncogenic and mitotic genes but notably still through almost entirely non-overlapping gene sets and genes. Subcomponent 3A, on the other hand, which is connected to hydrogen bonding and hydrophilic features, connects to a very different cellular response: the turning on of many stress-induced “defense” genes. In other words, we document how within the same compound or related compounds, hydrophobic and size features drive a mitotic arrest response while hydrogen bonding and hydrophilic features drive a reparative response. This knowledge, in combination with gene expression data in the solid tumors may allow us to design and utilize the chemotherapeutic agents with the appropriate balance of hydrophilic, size and hydrogen bonding for each cancer patient to hit the correct balance between anti-growths to damage response induction for best possible efficacy.

# Methods

## Gene Expression Data

We used the Connectivity Map gene expression profile data set as biological response profiles to drug treatments, forming the *biological space*. Instead of the rank-based procedure of the original Connectivity Map paper, we used a different preprocessing method since ranking amplifies noise. Even small differences in low intensities, which contain mostly noise are ranked, and this has a significant impact on the identification of differentially expressed genes. Hence, we downloaded the raw data files in original CEL-format, from http://www.broadinstitute.org/cmap/, which we RMA-normalized [21] before computing differential expression. We used expression profiles from the most abundant microarray platform (HT-HG-U133A) in the data collection. Differential expression was then taken with respect to the control measurements in each measurement batch. In the case of multiple controls per batch, we formed a more robust control by removing as an outlier the control with the highest Euclidean distance to the other controls, and then used the mean of the rest as the controls. To further reduce the noise in the expression data, we discarded 5% of the genes having the highest variance in the control measurements, that is, genes having high level of variation unrelated to chemical responses. As simple means of balancing between the varying sample sizes for different chemicals in the CMap data, we chose for each chemical the one instance with the strongest effect, measured with the highest norm of response, on the cell line for further analysis. The resulting gene expression data consisted of profiles for 1159 compounds over 11,350 genes.

To bring in prior knowledge of biological responses, and to reduce the dimensionality of the gene expression data, we performed Gene Set Enrichment Analysis (GSEA). GSEA gives as output for each gene set the direction (positive/negative) and strength of the activity, as measured by the false discovery rate (FDR) q-values, ranging from 0 (indicating highest activity) to 1. We transformed the q-values for the CCA by first inverting the q-values such that 1 indicates the highest activity, and then further mirroring q-values for the negatively activated gene sets with respect to zero to take the sign of activity into account. This results in a reasonably unimodal distribution of the data around zero, with higher positive and negative values indicating higher positive and negative activation of the gene sets, respectively. In the resulting data we have biological activation profiles over 1321 gene sets for 1159 distinct chemicals (see Figure 1.B).

As the gene sets, we used the C2 collection (curated gene sets v2.5) from the Molecular Signatures Database (http://www.broadinstitute.org/gsea/msigdb/). The extensive collection of gene sets covered 90% genes in our data and is thus a sensible representation of the biological space. GSEA was computed with the Java software package version 2-2.05 (http://www.broadinstitute.org/gsea).

## Chemical descriptors

The *chemical space* was formed by representing each chemical with a set of descriptors of its structure and function. In the analysis, the chemical similarity will be dependent on the selected descriptors and thus the selection is of utmost importance. This is especially true when the aim is to find small molecules that share targets and biological functions regardless of structural similarity. We use the VolSurf descriptors, calculated using MOE version 2009.10 (http://www.chemcomp.com/software.htm). Original sdf-files were translated into 3D using Maestro/LigPrep (Maestro version 9.0) since VolSurf descriptors are based on 3D molecular fields. The resulting data contains 76 descriptors for each chemical.

## Canonical Correlation Analysis

Drug action mechanisms are indirectly visible in relationships between the chemical properties of the drug molecules and the biological response profiles. We carry out a data-driven search for such relationships with a method that searches for correlated components in the two spaces, as shown in Figure 1.

Canonical Correlation Analysis (CCA [9]) is a multivariate statistical model for studying the interrelationships between two sets of variables. CCA explores correlations between the two spaces whose role in the analysis is strictly symmetric, whereas classical regression approaches like Partial Least Squares [22] typically explain one or possibly several response variables in one space by a set of independent variables in the other one. The result of the CCA analysis is an underlying component subspace relating chemical descriptors with gene sets.

Let us consider two matrices $X$ and $Y$, of the size of $n$ x $p$ and $n$ x $q$, representing the chemical and biological spaces. The rows represent the samples and the columns are the features (gene set activation values or chemical descriptors). In the following we describe the CCA learning algorithm as a stepwise process.

First, two projection vectors $w_1$ and $v_1$ are sought such that they maximize the correlation $P_1$ between components of data formed by projecting the data onto them,

$$P_1 = \max \mathrm{cor}_{w_1,v_1} \left( Xw_1, Yv_1 \right),$$

subject to the constraint that the variance of the components is normalized, *i.e.*, var$(Xw_1)$ = var$(Yv_1)$ = 1.

The resulting linear combinations $Xw_1$ and $Yv_1$ are called the first canonical variates or components, and $P_1$ is referred to as the first canonical correlation. The first canonical variates explain the maximum possible shared variance of the two spaces along a single linear pair of projections: $w_1$ and $v_1$.

The next canonical variates and correlations can be found as follows. For each successive step $s$ = 2,3,…min$(p,q)$, the projection vectors $(w_s,v_s)$, can be found by maximizing

$$P_s = \max \mathrm{cor}_{w_s,v_s} \left( Xw_s, Yv_s \right),$$

subject to the constraint var$(Xw_s)$ = var$(Yv_s)$ = 1, and with a further constraint of uncorrelatedness between different components.

Let $C_{xx} = XX^T$ and $C_{yy} = YY^T$ denote the scaled sample covariance matrices for the two input spaces, and $C_{xy} = XY^T$ the sample cross-covariance. Then the first canonical correlation is

$$P_1 = w_1^T C_{xy} v_1 \Big/ \sqrt{w_1^T C_{xx} w_1} \sqrt{v_1^T C_{yy} v_1} .$$

If $C_{xx}$ and $C_{yy}$ are invertible the vectors $w_1$ and $v_1$ maximizing the above equation can be found. Generally, in omics data and also in our study, the number of genes/gene sets is large compared to the number of experiments. In such cases the classical CCA solution may not exist or it can be very sensitive to collinearities among the variables. This issue can be addressed by introducing regularization [23-25], *i.e.*, by penalizing the norms of the associated vectors. Hence, we seek projection vectors that maximize the penalized correlation

$$P_1 = \frac{w_1^T C_{xy} v_1}{\sqrt{w_1^T C_{xx} w_1 + L_1 \|w_1\|} \sqrt{v_1^T C_{yy} v_1 + L_2 \|v_1\|}} .$$

The regularization coefficients $L_1$ and $L_2$ were estimated with 20-fold cross validation over a grid of values, while maximizing the retrieval performance on known drug properties. The retrieval procedure and performance measure are described in the drug similarity validation section below. In each fold, the model was first applied to a training data set, and the test data were then projected to the obtained components. Estimated regularization parameter values were $L_1$ = 100 and $L_2$ = 0.001. We used R-package “CCA” [24].

## Drug similarity validation procedure

To quantitatively validate the performance of the component model in extracting functionally similar drugs, we carried out the following analysis. For a given data set, we first computed pairwise similarities of drugs. In practice, we used each chemical in turn as a query, and ranked the other chemicals based on their similarity to the query. For the similarity measure, we had three alternatives, similarity in the CCA component space, in the biological space, and in the chemical space. Finally, we computed the average precision of retrieving chemicals that are functionally similar to the query, *i.e.* share at least one known property in an external validation set. We report the mean of the average precisions for all chemicals. We repeat the results as a function of the number of the top ranked chemicals used to compute the average precision (from 5 to 100).

We constructed the external validation set about the functional similarity of the drugs from their known protein targets and ATC (Anatomical Therapeutic Chemical, http://www.whocc.no/atc_ddd_index/) codes. Drug target information was obtained from ChEMBL (https://www.ebi.ac.uk/chembl/), DrugBank (http://www.drugbank.ca/), DUD (http://dud.docking.org/), and ZINC (http://zinc.docking.org/). We additionally extracted targets and ATC codes for the CMap chemicals from the supplementary material provided in [8]. From the ATC codes we used the fourth level information, indicating the chemical/therapeutic/pharmacological subgroup and hence high similarity of drugs. In total we have 4427 associations between 821 CMap chemicals and 796 targets or ATC codes.

## Visualizing through an “Eye diagram”: Relationship between gene sets, extracted components, and VolSurf descriptors

The CCA components summarize statistical relationships between the chemical and biological spaces. The relationships can be visualized as in Figure 1 and Figure 3: The components in the middle are connected to the chemical descriptors that activate them (left) and to the gene sets that are differentially expressed when the component is active. We selected the top 10 significant components from the CCA model for the visualizations. The significances of the components were estimated by a permutation test, using the observed correlations as a test statistic. The samples in one of the spaces were randomly rearranged removing the relationship with the other space. One thousand such random permutations were formed and their canonical correlations computed. Component significances were then determined as the proportion of random correlations that are greater than the observed correlation.

The components were further split into two subcomponents labeled ‘A’ and ‘B’, in A the canonical scores are positive and in B negative. Compounds in the two subcomponents behave in the opposite fashion on the gene sets and VolSurf features, such that when one of the subcomponents activates some biological processes, the other either has no effect or deactivates them. For visual clarity the eye diagram shows only the top 10 correlated gene sets for each component, out of the 1321 gene sets used. All 76 VolSurf features are shown. The eye diagram was originally introduced in [26] for visualizing component models.

## Differentially expressed genes and GO enrichment

To get a comprehensive view of the biology in each component we extracted the genes and Gene Ontology classes active in each one of them. For each component, we took the top 20 positively and top 20 negatively correlated gene sets and listed the genes in them. The differential expression of these genes was tested for activation/repression of the gene in the top 10 active compounds in the component using a regularized t-test [27]. The genes having p-values $< 0.05$ were considered to be significantly activated by the compounds in the component. This procedure ensures that the extracted genes are most representative of the top correlated gene sets in the component, hence contributing the most to the canonical correlation.

The component-specific list of significantly differentially expressed genes was used to compute the corresponding Gene Ontology Enrichment for each component. Enrichment was computed for Biological Process classes using GOstats R-package (www.bioconductor.org/help/bioc-views/release/bioc/html/GOstats.html).

## Characterizing drug response on breast cancer cells

We investigated if the components reveal interesting patterns in the responses to drugs, by plotting the transitions caused by each drug in the gene subspace defined by the component. This was done by extracting the 100 most significant genes as an effective representative of changes caused by treatments in the genome (using the procedure described in the above sub-section). The profiles of 30 independent cell lines in a steady-state, unperturbed conditions, were included to act as references for calibrating the display. These independent breast cancer cell lines were obtained from ArrayExpress experiment ID E-MTAB-37 (www.ebi.ac.uk/arrayexpress) with replicates merged to make a single representation for each of the cell types. All cell lines were annotated as BasalA, BasalB, Luminal, or progenitor using classifications by Kuemmerle et al.,[28]. Only MCF7 (breast cancer) treatments were used from CMap data.

The breast cancer cell line and CMap data come from different Affymetrix platforms, HG-U133plus_2.0 and HT-HG-U133A, respectively. We therefore normalized them separately by computing differential expression as the expression value divided by the mean of each gene within the platform. These normalized data were scaled using log2.

Both the CMap-selected instances and breast cancer cell data were collected into a single matrix. To visualize the transitions, pairwise correlation similarities were computed over this matrix and plotted using the state-of-the-art non-linear dimensionality reduction and visualization tool NeRV [29,30]. The result is a mapping of the high-dimensional expression profiles to a two-dimensional display for easier visualization, such that if two points are similar in the visualization, they can be trusted to have been similar before the projections also. NeRV visualization of component 3A, which is analyzed in the Results, is shown in Figure 5 .

# Acknowledgements

SuK, AF, JP, and SaK belong to the Adaptive Informatics Research Center at Aalto University School of Science. This work was supported by the Academy of Finland [Center of Excellence funding no. 213502 (Translational Genome-scale Biology) and funding no. 140057 (Computational modeling of the biological effects of chemicals)], the PASCAL2 Network of Excellence, ICT [216886], Sigrid Juselius Foundation, Cancer Society of Finland, the Jane and Aatos Erkko Foundation, the HBGS graduate school (J-PM), the FICS graduate school (SuK and AF), and the HECSE graduate school (JP).